\documentclass[twocolumn,superscriptaddress]{revtex4-1}
\usepackage{amsmath,amssymb,bm,graphicx,hyperref,color}
\allowdisplaybreaks[1]

\renewcommand\d{\partial}

\DeclareMathOperator\sgn{sgn}

\begin{document}

\title{Comparative study of one-dimensional Bose and Fermi gases with contact interactions\\
from the viewpoint of universal relations for correlation functions}

\author{Yuta Sekino}
\affiliation{Department of Physics, Tokyo Institute of Technology, Ookayama, Meguro, Tokyo 152-8551, Japan}
\author{Shina Tan}
\affiliation{School of Physics, Georgia Institute of Technology, Atlanta, Georgia 30332, USA}
\affiliation{Center for Cold Atom Physics, Chinese Academy of Sciences, Wuhan 430071, China}
\author{Yusuke Nishida}
\affiliation{Department of Physics, Tokyo Institute of Technology, Ookayama, Meguro, Tokyo 152-8551, Japan}

\date{October 2017}

\begin{abstract}
One-dimensional spinless Bose and Fermi gases with contact interactions have the close interrelation via Girardeau's Bose-Fermi mapping, leading to the correspondences in their energy spectra and thermodynamics.
However, correlation functions are in general not identical between these systems.
We derive in both systems the exact universal relations for correlation functions, which hold for any energy eigenstate and any statistical ensemble of the eigenstates with or without a trapping potential.
These relations include the large-momentum behaviors of static structure factors and of momentum distributions as well as energy relations, which connect the sums of kinetic and interaction energies to the momentum distributions.
The relations involve two- and three-body contacts, which are the integrals of local pair and triad correlations, respectively.
We clarify how the relations for bosons and fermions differ and are connected with each other.
In particular, we find that the three-body contact makes no contribution to the bosonic energy relation, but it plays a crucial role in the fermionic one.
In addition, we compute the exact momentum distribution for any number of fermions in the unitary limit.
\end{abstract}

\maketitle

\section{\label{sec:introduction}Introduction}
Understanding strongly correlated many-body systems is a challenging problem common to various fields in physics, such as atomic physics, condensed matter physics, and nuclear physics.
In ultracold atomic gases~\cite{Bloch:2008,Giorgini:2008,Inguscio:2007}, an interaction strength can be tuned by a Feshbach resonance~\cite{Chin:2010}.
By using an optical lattice, we can also obtain a low-dimensional gas, whose interaction is controlled by a confinement-induced resonance~\cite{Olshanii:1998,Granger:2004}.
This high controllability allows ultracold atoms to be ideal grounds to investigate quantum many-body systems with strong interactions not only in three dimensions (3D) but also in 2D and 1D.

When the interaction range of atoms is much smaller than other length scales such as a scattering length, a thermal de Broglie wavelength, and a mean interatomic distance, an atomic gas can be described by a model with a contact interaction.
In 1D, a system with a contact interaction has a peculiar property.
Bosons with an even-wave interaction have a one-to-one correspondence to spinless fermions with an odd-wave interaction~\cite{Granger:2004,Girardeau:2004}:
When a 1D scattering length of bosons $a_B^{e}$ and that of fermions $a_F^{o}$ are tuned to $a_B^{e}=a_F^{o}$, all energy eigenstates for bosons are mapped to those for fermions and vice versa via the Bose-Fermi mapping proposed by Girardeau~\cite{Girardeau:1960}.
This is a natural generalization of the well-known correspondence between impenetrable bosons and free fermions with $a_B^{e}=a_F^{o}=-0$~\cite{Girardeau:1960}.
As a consequence of this Bose-Fermi correspondence, the energy spectrum and thermodynamics are identical between bosons and fermions with $a_B^{e}=a_F^{o}$.
However, correlation functions for bosons are not always the same as those for fermions.

In homogeneous cases, 1D bosons and fermions with contact interactions are described by integrable models~\cite{Lieb:1963,Cheon:1999}.
Energy spectra and free energies in these systems can be exactly obtained by the Bethe ansatz~\cite{Lieb:1963,Yang:1969}.
On the other hand, the calculations of correlation functions are in general much more complicated even in integrable systems~\cite{Korepin:1993}.
One of the great successes in such a problem is Haldane's theory of quantum liquids, where the long-range scaling behaviors of correlation functions at zero temperature were derived~\cite{Haldane:1981}.
These universal behaviors originate from the quantum many-body fluctuation in 1D.

Recently, it has been elucidated that a system with a contact interaction has another type of universal property, which is effectively determined by few-body physics~\cite{Tan:2008,Braaten:2008,Braaten:2012}.
Here, various quantities including power-law tails of correlation functions at large momentum or high frequency, the energy, and the derivative of a free energy with respect to a coupling constant are related to so-called contact parameters, which measure short-range correlations of the system.
These universal relations are strong constraints on the system because they hold for any number of particles, strength of the interaction, temperature, and with or without a trapping potential.
While the relations were originally found in a 3D Fermi gas with an $s$-wave interaction~\cite{Tan:2008,Braaten:2008,Braaten:2012}, they have been generalized to various systems such as Bose gases~\cite{Braaten:2011,Werner:2012}, lower-dimensional gases~\cite{Werner:2012,Barth:2011,Hofmann:2012,Valiente:2011}, and quantum gases with higher partial-wave interactions~\cite{Yoshida:2015,Yu:2015,He:2016} as well as to nuclear systems~\cite{Weiss:2015}.
The relations have also been verified experimentally in 3D Fermi gases with $s$-wave and $p$-wave interactions~\cite{Stewart:2010,Luciuk:2016}.

In this paper, correlation functions for 1D bosons and fermions with contact interactions are studied from the viewpoint of universal relations.
In Sec.~\ref{sec:model}, we introduce the models of 1D bosons and fermions and review the Bose-Fermi mapping and its consequences.
For bosons and fermions, we derive the large-momentum tails of the static structure factor and the momentum distribution and apply them to uniform gases at zero temperature in Sec.~\ref{sec:tail}.
The energy relations, in which the sums of kinetic and interaction energies are expressed in terms of momentum distributions and contact parameters, are derived in Sec.~\ref{sec:energy}.
We clarify the consequences of the Bose-Fermi correspondence in these universal relations and verify them for homogeneous systems.
While we focus on energy eigenstates without a trap in the thermodynamic limit or those of trapped gases in Secs.~\ref{sec:model}--\ref{sec:energy}, the results therein are generalized to finite-size systems and statistical ensembles in Sec.~\ref{sec:generalization}.
In Sec.~\ref{sec:unitarity}, we exactly compute the momentum distribution for $N$ fermions in the unitary limit and demonstrate that the relations for fermions hold in this case.
We conclude this paper in Sec.~\ref{sec:conclusion}.

\section{\label{sec:model}Models}
We start with the first-quantized Hamiltonians for 1D bosons and fermions with a 1D scattering length, $a_B^{e}=a_F^{o}=a$.
The Hamiltonian for $N$ bosons with an even-wave interaction is given by
\begin{align}\label{eq:H_B}
H_B=H_0-\frac{2\hbar^2}{ma}\sum_{i<j}\delta(x_{ij}),
\end{align}
with
\begin{align}
H_0=\sum_{i=1}^{N}\left(-\frac{\hbar^2}{2m}\frac{\d^2}{\d x_i^2}+V(x_i)\right),
\end{align}
while the Hamiltonian for $N$ fermions with an odd-wave interaction is given by
\begin{align}\label{eq:H_F}
H_F=H_0-\frac{2\hbar^2a}{m}\sum_{i<j}\delta'(x_{ij})D_{ij}.
\end{align}
Here, $m$ is the mass of particles, $x_{ij}=x_i-x_j$ is the relative coordinate of the pair of particles $i$ and $j$, and the trapping potential $V(x)$ is an arbitrary smooth function of $x\in(-\infty,\infty)$.
The linear operator $D_{ij}$ acts on a fermionic wave function as $D_{ij}\Psi_F=\frac{\d}{\d x_{ij}}\Psi_F|_{x_{ij=+0}}$, with the center-of-mass coordinate $X_{ij}=(x_i+x_j)/2$ of the pair of fermions $i$ and $j$ and the other $N-2$ coordinates $\{x_k\}_{k\neq i,j}$ fixed.

The Bose-Fermi mapping is defined by
\begin{align}\label{eq:mapping}
\Psi_F(x_1,\dots,x_N)=A\Psi_B(x_1,\dots,x_N),
\end{align}
where $\Psi_B$ and $\Psi_F$ are bosonic and fermionic wave functions, respectively, the mapping factor
\begin{align}\label{eq:A}
A=A(x_1,\dots,x_N)=\prod_{i<j}\sgn(x_{ij})
\end{align}
is antisymmetric under the exchange of $x_i$ and $x_j$ with $i\neq j$, and the sign function $\sgn(x)$ is $+1$ for $x>0$ and $-1$ for $x<0$~\cite{Girardeau:1960}.
This mapping provides the one-to-one correspondence between the eigenstates of $H_B$ and $H_F$~\cite{Girardeau:2004,Cheon:1999}.
From the coupling constants in Eqs.~(\ref{eq:H_B}) and (\ref{eq:H_F}), one can see that weakly (strongly) interacting bosons correspond to strongly (weakly) interacting fermions.
Hereafter, we focus on a pair of bosonic and fermionic energy eigenstates, $\Psi_B$ and $\Psi_F$, related by Eq.~(\ref{eq:mapping}).

The existence of the mapping~(\ref{eq:mapping}) makes some physical quantities identical between $\Psi_B$ and $\Psi_F$~\cite{Girardeau:1960}.
Both states have the same energy $E$ and the same probability distribution of finding $M$ particles at the positions $x_1,\dots,x_M$:
\begin{align}\label{eq:g_M}
g_M(x_1,\dots,x_M)=\frac{N!}{(N-M)!}\int dx_{M+1}\cdots dx_N\nonumber\\
\times|\Psi_\alpha(x_1,\dots,x_N)|^2,
\end{align}
where $\alpha=B,F$ labels statistics of particles and $\Psi_\alpha$ is normalized as $\int dx_1\cdots dx_N|\Psi_\alpha|^2=1$.
Hereafter, we abbreviate the label $\alpha=B,F$ for quantities identical between $\Psi_B$ and $\Psi_F$.
Because of the correspondence of $g_M(x_1,\dots,x_M)$, the following quantities are also identical for $\Psi_B$ and $\Psi_F$:
the density profile $n(x)=g_1(x)$, the static structure factor
\begin{align}\label{eq:S(k)}
S(k)=1+\frac{1}{N}\int dx_1dx_2e^{-ik(x_1-x_2)}\nonumber\\
\times[g_2(x_1,x_2)-n(x_1)n(x_2)],
\end{align}
and the two- and three-body contacts,
\begin{align}\label{eq:contact}
C_2\equiv\int dx\,g_2(x,x),\qquad C_3\equiv\int dx\,g_3(x,x,x),
\end{align}
which are contact parameters given by the integrals of local pair and triad correlations, respectively.
The correspondence of the two-body contact $C_2$ between bosons and fermions was previously pointed out in Ref.~\cite{Cui:2016a}.
We note that $C_2$ in the literature is defined as the coefficient of the large-momentum tail of a momentum distribution (see, e.g., Refs.~\cite{Barth:2011,Cui:2016a,Valiente:2012}).
On the other hand, since the purpose of this paper is to make a comparison between 1D bosons and fermions, $C_2$ in Eq.~(\ref{eq:contact}) is defined so as to be identical between $\Psi_B$ and $\Psi_F$.
As a result, the tails of momentum distributions presented below in Eqs.~(\ref{eq:tail_of_rho_B}) and (\ref{eq:tail_of_rho_F}) may look different from those in the literature.

The correspondence of $g_M(x_1,\dots,x_M)$ for bosons and fermions is naturally generalized to systems at nonzero temperature $T$.
With $a$ and $T$ fixed, the canonical ensemble average of $g_M(x_1,\dots,x_M)$ is identical between bosons and fermions.
In particular, the two-body contact plays an important role in the thermodynamics of bosons and fermions.
By using the Hellmann-Feynman theorem, one can find that $C_2$ is the thermodynamic quantity conjugate to the inverse scattering length:
\begin{align}
\left(\frac{\d F}{\d(-1/a)}\right)_{T}=\frac{\hbar^2C_2}{m},
\end{align}
where $F$ is the free energy~\cite{Kheruntsyan:2003}.
For homogeneous gases in the thermodynamic limit, $C_2$ and $C_3$ were exactly computed in the bosonic Lieb-Liniger model by the Bethe ansatz~\cite{Kheruntsyan:2003,Gangardt:2003,Cheianov:2006,Kormos:2009,Kormos:2011}.

While $S(k)$ is a correlation function identical between $\Psi_B$ and $\Psi_F$, the momentum distribution
\begin{align}\label{eq:rho(k)}
\rho_\alpha(k)=N\int dx_2\cdots dx_N\left|\int dx_1e^{-ikx_1}\Psi_\alpha(x_1,\dots,x_N)\right|^2
\end{align}
is a correlation function strongly dependent on the statistics of particles.
However, in the next two sections, we show that there are two nontrivial connections between $\rho_B(k)$ and $\rho_F(k)$ resulting from the Bose-Fermi correspondence.

\section{\label{sec:tail}Tails of correlation functions}
Before evaluating the asymptotics of $S(k)$ and $\rho_\alpha(k)$ at large $k$, we recall the important property of contact interactions.
It is a well-known fact that contact interactions result in singular behaviors of energy eigenfunctions when two or more particles approach each other.
In particular, when two particles $i<j$ come close to each other, $\Psi_B$ and $\Psi_F$ satisfy the following boundary conditions:
\begin{subequations}\label{eq:short_range}
\begin{align}\label{eq:two_bosons}
\Psi_B&=(1-|x_{ij}|/a)\Phi_{B;ij}+O(x_{ij}^2),\\\label{eq:two_fermions}
\Psi_F&=[\sgn(x_{ij})-x_{ij}/a]\Phi_{F;ij}+O(x_{ij}^2),
\end{align}
\end{subequations}
where $\Phi_{\alpha;ij}=\Phi_{\alpha;ij}(X_{ij};\{x_k\}_{k\neq i,j})$.

We then derive the large-$k$ behavior of $S(k)$.
The key point to evaluate it is that the Fourier transform of a function having discontinuities or discontinuous derivatives in isolated points obeys a power law at $|k|\to\infty$.
The density profile $n(x)$ in Eq.~(\ref{eq:S(k)}) is a smooth function, so that its Fourier transformation rapidly vanishes for $|k|\to\infty$.
On the other hand, the singularity of $\Psi_\alpha$ in Eq.~(\ref{eq:short_range}) makes the pair-correlation function singular at short distance:
$g_2(x_1,x_2)=(1-2|x_{12}|/a)g_2(X_{12},X_{12})+O(x_{12}^2)$.
This singular term proportional to $|x_{12}|$ makes a dominant contribution to $S(k)$ in the large-$k$ limit.
By changing the integration variables $x_1,x_2\to x_{12}, X_{12}$, the large-$k$ behavior of $S(k)$ reads
\begin{align}
S(k)\simeq1-\frac{2}{Na}\int dx_{12} e^{-ikx_{12}} |x_{12}|C_2.
\end{align}
By using the formal Fourier transform of $|x|$, $\int dx e^{-ikx}|x|=-2/k^2$, we obtain the power-law tail of $S(k)$:
\begin{align}\label{eq:tail_of_S(k)}
S(k)\xrightarrow[|k|\to\infty]{}1+\frac{4C_2}{Nak^2}.
\end{align}
We note that this result is derived from the boundary conditions~(\ref{eq:short_range}) satisfied by all eigenstates of $H_\alpha$.
Therefore, this is a universal result in the sense that it holds for any eigenstate.

Let us now turn to the large-$k$ behavior of $\rho_\alpha(k)$.
In the case of bosons, the power-law tail of $\rho_B(k)$ at large $k$ was derived by Olshanii and Dunjko in a way similar to that for $S(k)$~\cite{Olshanii:2003}.
Their result is written as
\begin{align}\label{eq:tail_of_rho_B}
\rho_B(k)\xrightarrow[|k|\to\infty]{}\frac{4C_2}{a^2k^4}
\end{align}
in our definition of $C_2$.

We now derive the tail of $\rho_F(k)$.
The Fourier transform of $\Psi_F$ with respect to $x_1$ at large $k$ is dominated by the singularity in Eq.~(\ref{eq:two_fermions}):
\begin{align}
&\int dx_1e^{-ikx_1}\Psi_F (x_1,\dots,x_N)\nonumber\\
&\simeq\sum_{j=2}^Ne^{-ikx_j}\Phi_{F;1j}(x_j;\{x_k\}_{k\neq1,j})\int dx_{1j} e^{-ikx_{1j}}\sgn(x_{1j}),
\end{align}
where the change of variables $x_1\to x_{1j}$ was performed.
Since the Fourier transform of the sign function equals $\int dx e^{-ikx}\sgn(x)=-2i/k$, we find
\begin{align}
\int dx_1e^{-ikx_1}\Psi_F (x_1,\dots,x_N)\simeq\frac{-2i}{k}\sum_{j=2}^N e^{-ikx_j}\Phi_{F;1j}.
\end{align}
We then substitute this into Eq.~(\ref{eq:rho(k)}) and expand the modulus squared.
Because the cross terms are written as the Fourier transforms of functions continuous at $x_{jl}=0$, where $j,l\neq1$, with respect to $x_{jl}$,
they rapidly vanish in the large-$k$ limit.
By making use of $|\Phi_{F;1j}|=|\Psi_F|_{x_1=x_j}$ and the antisymmetry of $\Psi_F$, $\rho_F(k)$ is found to have the following power-law tail:
\begin{align}\label{eq:tail_of_rho_F}
\rho_F(k)\xrightarrow[|k|\to\infty]{}\frac{4C_2}{k^2}.
\end{align}
This is consistent with the result derived by using the operator product expansion~\cite{Cui:2016a}.

Let us compare Eq.~(\ref{eq:tail_of_rho_F}) with Eq.~(\ref{eq:tail_of_rho_B}).
As mentioned above, the mapping~(\ref{eq:mapping}) does not make $\rho_B(k)$ and $\rho_F(k)$ identical, and we indeed find that they obey different power laws at large momentum.
Nevertheless, Eqs.~(\ref{eq:tail_of_rho_B}) and (\ref{eq:tail_of_rho_F}) show that $\rho_B(k)$ and $\rho_F(k)$ are related to each other at $|k|\to\infty$ through the two-body contact: $\lim_{|k|\to\infty}a^2k^4\rho_B(k)=\lim_{|k|\to\infty}k^2\rho_F(k)=4C_2$.
This is one of the nontrivial connections between $\rho_B(k)$ and $\rho_F(k)$ resulting from the Bose-Fermi correspondence.

At the end of this section, we apply Eqs.~(\ref{eq:tail_of_S(k)}), (\ref{eq:tail_of_rho_B}), and (\ref{eq:tail_of_rho_F}) to the ground states of uniform Bose and Fermi gases with a negative scattering length.
In the thermodynamic limit, the states are characterized only by the dimensionless parameter $\gamma=-2/(na)>0$, where $n(x)=n$ is constant, and $C_2$ was exactly calculated for arbitrary $\gamma>0$~\cite{Gangardt:2003}.
This exact result combined with Eqs.~(\ref{eq:tail_of_S(k)}), (\ref{eq:tail_of_rho_B}), and (\ref{eq:tail_of_rho_F}) completely determines the large-$k$ asymptotics of $S(k)$ and $\rho_\alpha(k)$ for arbitrary $\gamma>0$.
For sufficiently small or large $\gamma$, the analytical expression of $C_2$ is also obtained as $C_2/N\simeq n$ for $\gamma\ll1$ and $C_2/N\simeq\frac{4}{3}\pi^2\gamma^{-2}n$ for $\gamma\gg1$.
By substituting these expressions into Eqs.~(\ref{eq:tail_of_S(k)}), (\ref{eq:tail_of_rho_B}), and (\ref{eq:tail_of_rho_F}), we obtain the explicit forms of the tails for $\gamma\ll1$ and $\gamma\gg1$, which are consistent with the large-$k$ limit of the previous results in Refs.~\cite{Bender:2005,Girardeau:2006,Cherny:2006}.

\section{\label{sec:energy}Energy relations}
We now derive the energy relations for 1D bosons and fermions.
The energy relation for bosons is easily derived by evaluating the expectation value of $H_B$ with respect to $\Psi_B$:
\begin{align}\label{eq:energy_bose}
E-E_{\mathrm{trap}}=\int\frac{dk}{2\pi}\frac{\hbar^2k^2}{2m}\rho_B(k)-\frac{\hbar^2C_2}{ma},
\end{align}
where $E_{\mathrm{trap}}=\int dx V(x)n(x)$ is a trapping energy~\cite{Valiente:2012}.
We note that the kinetic energy for bosons is ultraviolet convergent because of Eq.~(\ref{eq:tail_of_rho_B}).

In the case of fermions, an appropriate regularization procedure is however required to derive the energy relation.
If the expectation value of $H_F$ with respect to $\Psi_F$ was naively evaluated, one would be faced with divergences from both kinetic and interaction energies.
Indeed, Eq.~(\ref{eq:tail_of_rho_F}) shows that the kinetic energy has an ultraviolet divergence.
In this paper, we perform the regularization of the fermionic theory in the following way:
First, we introduce the function $f_\epsilon(x)$ with the range $\epsilon>0$, which is finite for $|x|<\epsilon$, rapidly vanishes for $|x|>\epsilon$, and approaches the derivative of the $\delta$ function in the zero-range limit, $\lim_{\epsilon\to0}f_\epsilon(x)=\delta'(x)$.
We then replace $\delta'(x)$ in Eq.~(\ref{eq:H_F}) with $f_\epsilon(x)$.
After evaluating the expectation value of the regularized $H_F$ with respect to $\Psi_F$, we take the zero-range limit $\epsilon\to0$.
Our approach is motivated by the method used in Ref.~\cite{Combescot:2009}.

Let us evaluate the expectation value of $H_0$ with respect to $\Psi_F$.
In the previous section, we show in the limit $\epsilon\to0$ that $\rho_F(k)$ behaves as $1/k^2$ for large momentum.
For $\epsilon>0$, Eq.~(\ref{eq:tail_of_rho_F}) holds as long as $|k|$ is much smaller than $1/\epsilon$ but much larger than the other momentum scales in the system.
However, the momentum distribution rapidly vanishes for $|k|>\Lambda\sim1/\epsilon$, which is easily demonstrated in the case of a simple finite-range potential such as a square-well potential.
As a result, a contribution from the region $|k|>\Lambda$ to the kinetic energy is negligible.
The expectation value of $H_0$ is thus found to be
\begin{align}\label{eq:kinetic_energy}
\int dx_1\cdots dx_N \Psi_F^*H_0\Psi_F\simeq\int_{-\Lambda}^\Lambda\frac{dk}{2\pi}\frac{\hbar^2k^2}{2m}\rho_F(k)+E_{\mathrm{trap}},
\end{align}
where the trapping energy $E_{\mathrm{trap}}=\int dx V(x)n(x)$ is convergent in the limit $\epsilon\to0$ and is the same as that for bosons.

The interaction energy has only two contributions remaining in the limit $\epsilon\to0$.
One contribution $U^{(2)}_F$ comes from the configuration where only one pair of fermions $i<j$ interact with each other.
In this region, the wave function behaves as Eq.~(\ref{eq:two_fermions}), leading to $-a D_{ij}\Psi_F=\Phi_{F;ij}$.
By taking the antisymmetry of $\Psi_F$ into account, $U^{(2)}_F$ can be written as
\begin{align}
&U^{(2)}_F=\frac{\hbar^2}{m}\int_{-\epsilon}^\epsilon dx_{12}f_\epsilon(x_{12})[\sgn(x_{12})-x_{12}/a]\nonumber\\
&\times\int_{-\infty}^\infty dX_{12}\int_{R_\epsilon} dx_3\cdots dx_N N(N-1)|\Phi_{F;12}|^2
+O(\epsilon).
\end{align}
Here, the symbol $R_\epsilon$ refers to the region where all fermions at $x_3,\dots,x_N$ are not affected by the interaction, i.e., $|x_{1i}|,|x_{2i}|>\epsilon$ for $i=3,\dots,N$ and $|x_{jk}|>\epsilon$ for $3\leq j<k\leq N$.
Because power counting with respect to $\epsilon$ provides $dx_{12},\, x_{12}=O(\epsilon)$, $f_\epsilon(x_{12})=O(\epsilon^{-2})$, and $X_{12},\,\sgn(x_{12})=O(1)$, we need to evaluate the integration over $R_\epsilon$ up to $O(\epsilon^2)$.
It is convenient to rewrite this integral as follows:
\begin{align}
&\int_{R_\epsilon} dx_3\cdots dx_N N(N-1)|\Phi_{F;12}|^2\nonumber\\
&=\left(\int_{\mathbb{R}^{N-2}}-\int_{\bar{R}_\epsilon}\right)dx_3\cdots dx_N N(N-1)|\Phi_{F;12}|^2,
\end{align}
where $\bar{R}_\epsilon$ denotes the complement of $R_\epsilon$.
The first integral equals $g_2(X_{12},X_{12})$.
The second is dominated by the region where only one of the remaining fermions approaches the pair of fermions 1 and 2, so that it is evaluated as $-(2\epsilon+|x_{12}|)g_3(X_{12},X_{12},X_{12})+O(\epsilon^2)$.
After performing the integration over $x_{12}$ and $X_{12}$, $U^{(2)}_F$ is found to be
\begin{align}\label{eq:U^{(2)}_F}
U^{(2)}_F=\frac{\hbar^2}{m}(J_\epsilon C_2-2\epsilon J_\epsilon C_3+C_2/a+C_3)+O(\epsilon),
\end{align}
where $J_\epsilon\equiv\int_{-\epsilon}^\epsilon dx_{12}f_\epsilon(x_{12})\sgn(x_{12})$ and the three-body contact $C_3$ is defined in Eq.~(\ref{eq:contact}).
By using the Fourier transforms of $\delta'(x)$ and $\sgn(x)$ and recalling the existence of the momentum cutoff $\Lambda$, $J_\epsilon$ is evaluated as $J_\epsilon\simeq-2\int_{-\Lambda}^\Lambda dk/(2\pi)$.
The first term in Eq.~(\ref{eq:U^{(2)}_F}) thus cancels the divergence from the kinetic energy in Eq.~(\ref{eq:kinetic_energy}).

The other contribution $U^{(3)}_F$ to the interaction energy comes from the configuration where only three fermions interact with each other.
When three fermions at $x_1$, $x_2$, and $x_3$ come close to each other, the Bose-Fermi mapping~(\ref{eq:mapping}) implies that $\Psi_F^*$ has the singularity $\sim\sgn(x_{12})\sgn(x_{13})\sgn(x_{23})=\sgn(x_{12}x_{13}x_{23})$.
By taking the antisymmetry of $\Psi_F$ into account, this contribution can be evaluated as
\begin{align}
U^{(3)}_F=\frac{\hbar^2}{m}\int_{-\epsilon}^\epsilon dx_{12}\int_{-\infty}^\infty dX_{12} f_\epsilon(x_{12}) g_3(X_{12},X_{12},X_{12})\nonumber\\
\times\int_{x_\mathrm{min}-\epsilon}^{x_\mathrm{max}+\epsilon} dx_3\sgn(x_{12}x_{13}x_{23})
+O(\epsilon),
\end{align}
where $x_\mathrm{min}=\mathrm{min}(x_1,x_2)$ and $x_\mathrm{max}=\mathrm{max}(x_1,x_2)$.
This integration can be easily performed to obtain
\begin{align}\label{eq:U^{(3)}_F}
U^{(3)}_F=\frac{\hbar^2}{m}(2\epsilon J_\epsilon+1)C_3+O(\epsilon).
\end{align}

Finally, by summing up Eqs.~(\ref{eq:kinetic_energy}), (\ref{eq:U^{(2)}_F}), and (\ref{eq:U^{(3)}_F}) and taking the limit of $\epsilon\sim1/\Lambda\to0$, we arrive at the following energy relation:
\begin{align}\label{eq:energy_fermi}
E-E_{\mathrm{trap}}&=\lim_{\Lambda\to\infty}\int_{-\Lambda}^\Lambda\frac{dk}{2\pi}\frac{\hbar^2k^2}{2m}\left(\rho_F(k)-\frac{4C_2}{k^2}\right)\nonumber\\
&\quad+\frac{\hbar^2C_2}{ma}+\frac{2\hbar^2C_3}{m}.
\end{align}
This is a novel universal relation including the three-body correlation in 1D.
We note that Eq.~(\ref{eq:energy_fermi}) is similar to the energy relation for 3D bosons, where the Efimov effect takes place~\cite{Efimov:1970}, in the sense that both of them involve the three-body contacts~\cite{Braaten:2011}.

The energy relation for 1D fermions with an odd-wave interaction was recently proposed in Ref.~\cite{Cui:2016a}, where the three-body contribution to the relation was not included.
To demonstrate the necessity of this contribution, we apply Eq.~(\ref{eq:energy_fermi}) to a uniform Fermi gas in the thermodynamic and unitary limits, $a\to-\infty$, at zero temperature.
In this case, the Fermi gas corresponds to the ideal Bose gas and has $E=E_{\mathrm{trap}}=0$, $C_2/N=n$, and $C_3/N=n^2$, with $n(x)=n$.
The momentum distribution, $\rho_F(k)/N=4n/(k^2+4n^2)$, was also exactly calculated~\cite{Bender:2005,Girardeau:2006}.
By substituting these into both sides of Eq.~(\ref{eq:energy_fermi}), one can see that Eq.~(\ref{eq:energy_fermi}) holds in this case and that the three-body contact makes an essential contribution to the energy of fermions.
On the other hand, the virial theorems for bosons and fermions trapped by a harmonic potential do not involve $C_3$ and are identical to each other~\cite{Cui:2016a,Valiente:2012}.

We now compare the energy relations for bosons and fermions.
Although the left-hand sides of Eqs.~(\ref{eq:energy_bose}) and (\ref{eq:energy_fermi}) are the same, the right-hand sides look quite different, in particular, in the absence or presence of a term proportional to $C_3$.
Nevertheless, Eqs.~(\ref{eq:energy_bose}) and (\ref{eq:energy_fermi}) connect $\rho_B(k)$ and $\rho_F(k)$, which is the other nontrivial connection of $\rho_\alpha(k)$ resulting from the Bose-Fermi correspondence.

\section{\label{sec:generalization}Generalization of results}
The universal relations [Eqs.~(\ref{eq:tail_of_S(k)}), (\ref{eq:tail_of_rho_B}), (\ref{eq:tail_of_rho_F}), (\ref{eq:energy_bose}), and (\ref{eq:energy_fermi})] presented in the previous sections can be generalized to bosons and fermions with a finite system size $L<\infty$.
Although $k$ is not continuous but quantized, the power-law tails of correlation functions [Eqs.~(\ref{eq:tail_of_S(k)}), (\ref{eq:tail_of_rho_B}), and (\ref{eq:tail_of_rho_F})] hold for these systems.
The energy relations for these systems can be obtained by replacing the integrals in Eqs.~(\ref{eq:energy_bose}) and (\ref{eq:energy_fermi}) with the sum over $k$: $\int_{-\Lambda}^\Lambda dk/(2\pi)\to L^{-1}\sum_{|k|<\Lambda}$.
In the next section, we demonstrate Eqs.~(\ref{eq:tail_of_rho_F}) and (\ref{eq:energy_fermi}) for a finite-size fermionic system at unitarity $a\to\infty$.

Here, we note a subtle issue in the Bose-Fermi mapping~(\ref{eq:mapping}) with periodic (or more generally twisted) boundary conditions~\cite{Lieb:1963,Cheon:1999}:
If and only if $N$ is even, the factor $A$ defined by Eq.~(\ref{eq:A}) makes boundary conditions for the mapped state opposite in sign from those for the original state.
For example, $\Psi_B$ with periodic boundary conditions corresponds to $\Psi_F$ with antiperiodic (periodic) ones if $N$ is even (odd).
On the other hand, the factor $A$ does not change hard wall boundary conditions~\cite{Hao:2006,Hao:2007}.

Since the universal relations hold for any energy eigenstates, any statistical ensemble of the bosonic eigenstates satisfies Eqs.~(\ref{eq:tail_of_S(k)}), (\ref{eq:tail_of_rho_B}), and (\ref{eq:energy_bose}), while any statistical ensemble of the fermionic eigenstates satisfies Eqs.~(\ref{eq:tail_of_S(k)}), (\ref{eq:tail_of_rho_F}), and (\ref{eq:energy_fermi}).
This set of the ensembles includes not only a canonical ensemble in thermal equilibrium but also a generalized Gibbs ensemble, which is expected to describe the stationary properties after a quantum quench in integrable systems~\cite{Rigol:2007}.

\section{\label{sec:unitarity}Fermions at unitarity}
In this section, we study the ground state of a finite number of spinless fermions in the unitary limit $a\to\infty$.
For simplicity, we consider fermions which correspond to free bosons with periodic boundary conditions without an external potential:
\begin{align}\label{eq:unitary}
\Psi_F(x_1,\dots,x_N)= L^{-N/2} A(x_1,\dots,x_N),
\end{align}
where $0\leq x_i<L$.
This state has $E=E_{\mathrm{trap}}=0$, $C_2=N(N-1)/L$, and $C_3=N(N-1)(N-2)/L^2$.

We now compute $\rho_F(k)$ for the ground state.
Because the wave function in Eq.~(\ref{eq:unitary}) satisfies periodic (antiperiodic) boundary conditions if $N$ is odd (even), $k$ is quantized as $k=2n\pi /L$ [$k=(2n+1)\pi /L$], where $n$ is an integer.
After a lengthy but straightforward calculation, $\rho_F(k)$ is exactly obtained for any $N\geq2$:
\begin{align}\label{eq:rho_unitary}
\frac{\rho_F(k)}{L}=\delta_{k,0}+ (1-\delta_{k,0})\sum_{l=1}^{\lfloor N/2 \rfloor}\frac{(-1)^{l-1}N!}{(N-2l)!}\left( \frac{2}{kL} \right)^{2l},
\end{align}
where the floor function $\lfloor N/2\rfloor$ equals $(N-1)/2$ if $N$ is odd and $N/2$ if $N$ is even.
By taking the large-$k$ limit, we have $\rho_F(k)\to 4N(N-1)/(k^2L)$, which is consistent with Eq.~(\ref{eq:tail_of_rho_F}).
In addition, we confirmed that the energy relation for a finite-size system holds,
\begin{align}
\frac{1}{L}\sum_k\frac{\hbar^2k^2}{2m}\left(\rho_F(k)-\frac{4C_2}{k^2}\right)+\frac{2\hbar^2C_3}{m}=0,
\end{align}
by using \textsc{mathematica} at least up to $N=200$.

\begin{table*}[t]\renewcommand\arraystretch{1.6}
\caption{\label{tab:relations}
Comparison of universal relations for 1D bosons and fermions with contact interactions.
Here, $E$, $E_\mathrm{trap}$, $C_2$, $C_3$, and $S(k)$ are identical between bosons and fermions at the same scattering length $a$, but $\rho_B(k)$ and $\rho_F(k)$ differ.}\smallskip
\begin{ruledtabular}
\begin{tabular}{cccc}
Bosons&&Fermions&\\\hline
\multicolumn{3}{c}{$S(k)\xrightarrow[|k|\to\infty]{}1+\frac{4C_2}{Nak^2}$}&(\ref{eq:tail_of_S(k)})\\
$\rho_B(k)\xrightarrow[|k|\to\infty]{}\frac{4C_2}{a^2k^4}$&(\ref{eq:tail_of_rho_B})
&$\rho_F(k)\xrightarrow[|k|\to\infty]{}\frac{4C_2}{k^2}$&(\ref{eq:tail_of_rho_F})\\
$E-E_{\mathrm{trap}}=\int\frac{dk}{2\pi}\frac{\hbar^2k^2}{2m}\rho_B(k)-\frac{\hbar^2C_2}{ma}$&(\ref{eq:energy_bose})
&$E-E_{\mathrm{trap}}=\lim_{\Lambda\to\infty}\int_{-\Lambda}^\Lambda\frac{dk}{2\pi}\frac{\hbar^2k^2}{2m}\left(\rho_F(k)-\frac{4C_2}{k^2}\right)+\frac{\hbar^2C_2}{ma}+\frac{2\hbar^2C_3}{m}$&(\ref{eq:energy_fermi})
\end{tabular}
\end{ruledtabular}
\end{table*}

\section{\label{sec:conclusion}Conclusion}
We studied 1D bosons and fermions with contact interactions which are related to each other through the Bose-Fermi mapping~(\ref{eq:mapping}) and derived the power-law tails of $S(k)$ and $\rho_\alpha(k)$ at large $k$ and the energy relations, which are summarized in Table~\ref{tab:relations}.
We found the following three facts in these universal relations:
$S(k)$ has an identical tail between bosons and fermions;
the Bose-Fermi correspondence results in two nontrivial connections between $\rho_B(k)$ and $\rho_F(k)$ through their tails and through the energy relations;
and the three-body contact makes no contribution to the energy relation for bosons, but it makes an essential contribution to that for fermions.
Furthermore, Eqs.~(\ref{eq:tail_of_S(k)}), (\ref{eq:tail_of_rho_B}), and (\ref{eq:tail_of_rho_F}) together with the Bethe ansatz completely determine the large-$k$ tails of $S(k)$ and $\rho_\alpha(k)$ for uniform Bose and Fermi gases at any temperature.
We also computed $\rho_F(k)$ for the ground state of $N$ fermions in the unitary limit $a\to\infty$ and confirmed Eqs.~(\ref{eq:tail_of_rho_F}) and (\ref{eq:energy_fermi}) in this case.

We can consider some applications and generalizations of the universal relations presented in this paper.
Our relations can be used as reliable tests on numerical studies of correlation functions~\cite{Astrakharchik:2003,Caux:2006,Caux:2007,Jacqmin:2012,Panfil:2014}.
One may compute higher-order corrections to Eqs.~(\ref{eq:tail_of_S(k)}), (\ref{eq:tail_of_rho_B}), and (\ref{eq:tail_of_rho_F}) at large momentum in a way similar to 2D and 3D cases~\cite{Braaten:2011,Bellotti:2013}.
It should be interesting to see whether and how $C_3$ appears in these corrections.
One can also generalize energy relations to 1D bosons and fermions with finite effective ranges~\cite{Imambekov:2010,Qi:2013,Cui:2016b}, where a three-body correlation will make an essential contribution in the fermionic case as in Eq.~(\ref{eq:energy_fermi}).

\acknowledgments
The authors thank Shun Uchino and Takato Yoshimura for stimulating discussions.
The work by Y.\,S.\ and Y.\,N.\ was supported by JSPS KAKENHI Grants No.\ JP15K17727 and No.\ JP15H05855,
while the work by S.\,T.\ was supported by the US National Science Foundation CAREER award under Grant No.\ PHY-1352208.

\end{document}